\documentclass[
  ,draft            
  ]
  {aipproc}

\layoutstyle{6x9}

\usepackage{epsfig}
\usepackage{slashed}
\def\Journal#1#2#3#4{{#1} {\bf #2}, #3 (#4)}

\def\NCS{\em Nuovo Cimento Suppl.}

\def\AP{{\em Ann. Phys.}}
\def\NPA{{\em Nucl. Phys.} A}

\def\PRL{\em Phys. Rev. Lett.}
\def\PR{\em Phys. Rev.}
\def\PRC{{\em Phys. Rev.} C}
\def\PRD{{\em Phys. Rev.} D}

\def\be{\begin{equation}}
\def\ee{\end{equation}}
\def\bea{\begin{eqnarray}}
\def\eea{\end{eqnarray}}

\begin{document}

\title{Spectator theory for three-nucleon electromagnetic currents}

\author{Alfred Stadler}{
  address={Departamento de F\'isica, Universidade de \'Evora, \'Evora, Portugal, and}
  ,address={Centro de F\'isica Nuclear da Universidade de Lisboa, Lisboa, 
  Portugal} 
}

\begin{abstract} 
In this talk I present some of the more recent developments within the
Covariant Spectator Theory. My focus will be on aspects of the derivation of
gauge invariant electromagnetic three-nucleon currents which are consistent
with the hadronic equations and with the basic assumptions of this framework.
Important dynamical ingredients of the three-nucleon currents are also
discussed, namely the three-nucleon bound state vertex function and the
two-nucleon interaction model they are derived from. 

\end{abstract}

\maketitle


\section{The Covariant Spectator Theory}

The Covariant Spectator Theory (CST) was introduced already several decades ago by Franz
Gross \citep{Gro69}. Its purpose is to include relativity in a manifestly covariant way in
few-body problems, while keeping their complexity at a level that is still manageable for
numerical calculations. It has been already applied successfully to the description of a
variety of systems, including the deuteron and nucleon-nucleon ($NN$) scattering
\citep{Buc79, Gro92},
 elastic and inelastic electron scattering off the
deuteron \citep{Arn80, Ord95, Ada02}, and the 3N bound state \citep{Sta97, Sta97b}. 

The CST is based on Relativistic Quantum Field Theory, where an exact and complete
scattering amplitude can be expressed in terms of the Bethe-Salpeter equation (BSE)
\citep{Sal51}. The Spectator Equation (SE) is a particular re-organization of the complete
BSE for a system of heavy particles (nucleons) that interact via the exchange of lighter
particles (mesons). Ignoring vertex and self-energy corrections as well as genuine many-body
forces, the scattering amplitude is given as an infinite sum of ladder and crossed ladder
diagrams of all orders in the coupling constants. It can be generated by an integral
equation if the propagators and the irreducible kernel are chosen in a suitable way. 

While the BSE includes the full off-shell propagators for all heavy particles, the SE
restricts all but one of them to their mass shells. Consequently, contributions to the
scattering amplitude which are generated by one equation through iterations may appear as
part of the irreducible kernel of the other. When their respective kernels are
truncated---which in practical calculations is unavoidable since the complete kernels
themselves contain already an infinite number of diagrams---the BSE and the SE are no
longer equivalent. It has been shown that the truncated SE, in certain circumstances,
converges faster to the full result than the truncated BSE \citep{Gro82}. The simplest and
most often used cases are truncations at lowest order, leading to the so-called ladder
approximation since crossed-ladder graphs are thereby excluded. 

The SE has sometimes been misunderstood as an approximation to the BSE. This point of view
is somewhat misleading, since there is really nothing more fundamental about the BSE than
the SE (or other so-called quasi-potential equations). It is less confusing to understand
the SE as an equation in its own right. It is meant to replace the Schr\"odinger equation
in situations where relativity is important, rather than to approximate a truncated BSE. 

Some of the important features of the SE are: it is manifestly covariant although the loop
integrations are three-dimensional, all boosts are kinematic ({\it i.e.}, interaction
independent), the off-shell particle has negative-energy components, and cluster
separability holds. The latter property guarantees that the solution of the two-body SE is
consistent with the one that appears as dynamical input in the three-body SE.

\section{Two-nucleon scattering}

\begin{figure}
\includegraphics[width=6cm]{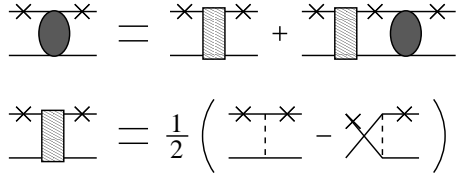}
\caption{Two-nucleon SE (upper panel) with antisymmetrized kernel (lower
panel). The symbol ``$\times$'' on a nucleon line indicates that the particle is
restricted to its positive-energy mass shell.} \label{F:2Ngrosseq}
%
\end{figure}

The SE for $NN$ scattering is shown in diagrammatic form in Fig.\ \ref{F:2Ngrosseq}. Its
antisymmetrized kernel (or ``potential'') is truncated to the lowest-order ladder terms,
such that it is of one-boson exchange (OBE) form. 

The first realistic OBE $NN$ potentials for the SE were constructed by Gross,
Van Orden, and Holinde \citep{Gro92}. They are based on the exchange of six
different mesons, two pseudoscalar ($\pi$ and $\eta$), two scalar ($\sigma$ and
$\delta$), and two vector mesons ($\rho$ and $\omega$). 

The potentials have meson-nucleon vertices
of the following form:
\be
g_s \Lambda_s = g_s \left[ 
 1 + \frac{\nu_s}{2m} ( \slashed{p}'- m + \slashed{p} - m ) 
 + \frac{\kappa_s}{4m^2} (\slashed{p}' - m)(\slashed{p} - m) \right] \, ,
\label{eq:Vs}
\ee
for scalars,
\bea
g_p \Lambda_p & = &
g_p \left[ \gamma^5 +
          \frac{\nu_p}{2m}\left[ (\slashed{p}' -m)\gamma^5
                               + \gamma^5 (\slashed{p} -m) \right]
       + \frac{\kappa_p}{4m^2}(\slashed{p}'-m)\gamma^5(\slashed{p}-m)\right]
\nonumber \\
& = & g_p \left[ (1-\nu_p)\gamma^5
     + \frac{\nu_p}{2m}\gamma^5 \slashed{q}
     + \frac{\kappa_p}{4m^2}(\slashed{p}'-m)\gamma^5(\slashed{p}-m)\right] \, ,
\label{eq:Vp}
\eea
for pseudoscalars, and
\be
g_v \Lambda_v^\mu =
g_v \left[ \gamma^\mu +
     \frac{\kappa_v}{2m} i \sigma^{\mu\nu} q_\nu +
     \frac{\kappa_v (1-\lambda_v)}{2m}
     \left[(\slashed{p}'-m)\gamma^\mu + \gamma^\mu(\slashed{p}-m)\right]
     + \cdots \right]
\ee
for vector mesons. Here, $p$ and $p'$ are the nucleon four-momenta, and $q$  is
the meson four-momentum at the meson-$NN$ vertex. Furthermore, $m$ is the
nucleon mass, the $g$'s are coupling constants, $\kappa_v$ is the usual $f/g$
ratio for vector mesons,  $\nu_s$, $\nu_p$, $\kappa_s$, $\kappa_p$,  and
$\lambda_v$ are off-shell coupling constants. The vertices are regularized by
form factors, one for each particle at the vertex, which are parametrized by
cut-off masses. 

While $\nu_p$ has appeared before as mixing parameter in potentials that allow 
a mixing of pseudoscalar and pseudovector $\pi NN$ coupling, the scalar
off-shell coupling proportional to $\nu_s$ is a new feature not explored
previously in $NN$ scattering. It contributes only if at least one of the
nucleons at the vertex is off mass shell. In frameworks that do not contain
off-shell nucleons, such as in relativistic hamiltonian dynamics or in
nonrelativistic quantum mechanics, these couplings vanish.

The terms proportional to $\kappa_s$ and $\kappa_p$ contribute only if both
nucleons at the vertex are off mass shell. With their inclusion, 
Eqs.~(\ref{eq:Vs}) and (\ref{eq:Vp}) represent the most general coupling
of nucleons to scalar or pseudoscalar mesons, respectively. However, in the
potentials fitted so far we have always set $\kappa_s = \kappa_p = 0$.

\section{The three-nucleon bound state}

The SE for the $3N$ bound state has been derived \citep{Sta97b} in complete
consistency with the CST of $NN$ scattering. This is achieved by
restricting in all intermediate states spectator particles to their
positive-energy mass shell (hence the name ``spectator theory''). This simple
prescription automatically insures that in any intermediate state only one
nucleon is off mass shell. As a consequence, loop integrations are again reduced
from four to three dimensions, the same as in the nonrelativistic case. 

\begin{figure}
\includegraphics[width=6cm]{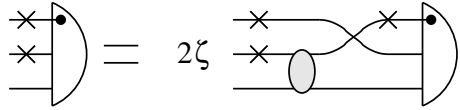}
\caption{The SE for the vertex function of the $3N$ bound state. The oval
represents the $NN$ spectator amplitude calculated from the SE of Fig.\
\ref{F:2Ngrosseq}. The dot in the vertex function indicates the spectator
particle during the last two-body interaction.} 
\label{F:3Nbound}
\end{figure}
Figure \ref{F:3Nbound} displays the homogeneous integral equation for the $3N$
vertex function in graphical form. It is given here in its general form, valid
both for
identical fermions ($\zeta=-1$) and bosons ($\zeta=+1$).

The numerical solution of the Faddeev-type $3N$ SE was performed \citep{Sta97} in a basis of
partial wave helicity states. The propagator of the off-shell nucleon is
decomposed into positive and negative energy components, or $\rho$-spin states,
which also enter into the classification of the basis states.

The $3N$ SE was solved for a family of $NN$ potentials, which were constructed
by fixing the scalar off-shell coupling parameters to a particular value and
fitting the remaining parameters to the energy-dependent \mbox{Nijmegen} $np$ phase
shift analysis \citep{Sto93} of 1993. Then the $\chi^2$ to $NN$ databases is
calculated from the potentials in a second step, using Arndt's code
SAID \citep{ArnSAID}. While the off-shell couplings for the $\sigma$ and $\delta$
mesons are in principle independent, initially we related them in terms of a common scaling
parameter $\nu$,  such that $\nu_\sigma = 2.6 \nu$ and $\nu_\delta = -0.75 \nu$
(a result obtained in preliminary independent fits). The parameters of some of
these potentials can be found in Ref.\ \citep{Sta98}.

It turned out that potentials with nonzero values of $\nu$ have a lower $\chi^2$ and
are therefore preferred by the fits. The potential named W16 has the minimum value,
$\chi^2=1.895$, for $\nu=1.6$, which corresponds to $\nu_\sigma=4.16$ and
$\nu_\delta=-1.2$. The triton binding energy varies rather strongly with $\nu$. The best
model in terms of $\chi^2$, W16, yields with $E_t=-8.491$ MeV also the binding energy
closest to the experimental value of $-8.48$ MeV.

This family of $NN$ potentials was designed principally to study the scalar off-shell
coupling and other relativistic effects in the $3N$ bound state. The number of adjustable
parameters was---with only 13---kept comparatively small. We are now in the process of
further improving the fit to the $NN$ data by allowing more potential parameters to vary
independently. As an example I mention here the (preliminary) model WJL19-1.1 with 19 free
parameters. Among other new features, it has different masses for the charged and neutral
pions, and its scalar off-shell coupling parameters, $\nu_\sigma=-6.59$ and
$\nu_\delta=2.67$, were varied independently. With a much improved $\chi^2=1.254$ it
demonstrates that the potentials of the CST are comparable in quality with the best
nonrelativistic potentials. They are, however, true meson-exchange potentials in the sense
that their parameters are the same in all partial waves or isospin channels. The model
WJL19-1.1 yields a somewhat higher triton binding energy of $E_t=-9.116$ MeV. This value is
likely to change as the fit is being finalized.

We also calculated the $3N$ bound state in the nonrelativistic limit, using a
potential fitted especially for this case. This calculation yields a $3N$
binding energy of -7.914 MeV. One can use this case also to estimate the
importance of relativistic boost effects, which here turned out to be repulsive
but relatively small.

\section{Spectator theory of electromagnetic three-nucleon currents}

For the calculation of elastic and inelastic electron scattering from the $3N$
bound state a conserved electromagnetic $3N$ current has to be derived within
the CST. As we will see, special care is needed to avoid double
counting and to arrive at expressions consistent with the basic assumptions of
the spectator formalism.

In the one-photon approximation, a gauge invariant current is obtained if all Feynman
diagrams are summed in which the photon is attached to all propagators and to all vertices
with momentum-dependent couplings. This is a clear prescription for any given Feynman
diagram. However, if we are dealing with an infinite sum of diagrams, defined through an
integral equation, a more general procedure is needed that systematically attaches the
photon in all the right places. 

\begin{figure}
\epsfig{file=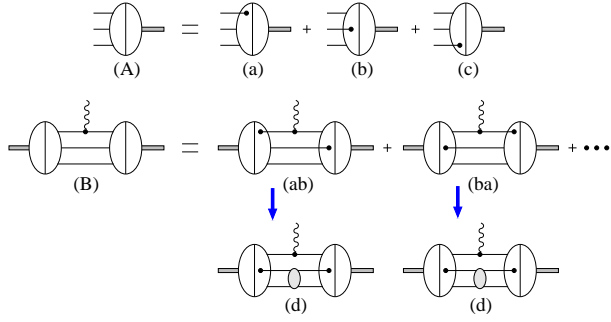,width=8cm}
\caption{An example for the overcounting problem in the BSE. The upper part
displays the decomposition of the full vertex function into components with a
given particle, indicated by a dot, as spectator during the last $NN$
interaction in the final state. The lower part shows how certain
contributions to the impulse approximation current appear twice if one simply
takes the ``obvious'' diagram (B) from nonrelativistic theory.} \label{F:overcounting}
%
\end{figure}

Unfortunately, one cannot simply use results from the nonrelativistic theory as
guidance. It is instructive to consider the elastic current in impuls
approximation for the BSE \citep{Kvi99}. Figure \ref{F:overcounting} shows that
diagram \ref{F:overcounting}(B), which one might naivly identify with the
impulse approximation in analogy with the nonrelativistic case, includes some
contributions twice, for instance diagram \ref{F:overcounting}(d). 

The origin of this double-counting problem can be traced back to the fact that
the two-body amplitude in diagram \ref{F:overcounting}(d) can be ``pulled
out'', by applying the $3N$ equation, from the vertex function in the initial as
well as in the final state. In contrast to nonrelativistic diagrams, in a
Feynman diagram it makes no difference if this two-body amplitude is extracted
from the right or left side of the diagram, since it can ``slide freely'' past the
point were the photon is attached to the third nucleon.
The double-counting problem in covariant scattering theories has been known for a long
time, especially in the context of simple systems of nucleons and pions \citep{Tay63,
Phi95}.

\begin{figure}[b]
\epsfig{file=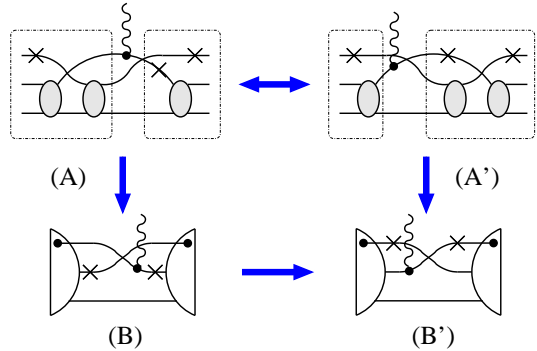,width=7cm}
\caption{An example of how diagrams with spectator particles off mass shell are
related to others with spectators on mass shell. See the discussion in the text.} \label{F:equivalence}
%
\end{figure}

A method to generate systematically the infinite series of diagrams that
represents the current, and avoids double counting, was introduced by
Kvinikhidze and Blankleider \citep{Kvi97a}. They call it the ``gauging of integral
equations method.'' It is based on the observation that coupling a photon to an
amplitude given by an integral equation satisfies the same distributive rule as
the differentiation of a product. For instance, if 
\be
\Gamma = 2 M G P \Gamma 
\ee
is the Faddeev equation for the vertex function $\Gamma$, with $M$ being the
$NN$ amplitude, $G$ an off-shell propagator, and $P$ the permutation
operator interchanging particles 1 and 2, then the gauged vertex function
satisfies
\be
\Gamma^\mu = ( 2 M G P \Gamma)^\mu 
=
2 M^\mu G P \Gamma + 2 M G^\mu P \Gamma 
+ 2 M G P \Gamma^\mu \, .
\ee
Kvinikhidze and Blankleider applied their gauging method also to the $3N$
SE \citep{Kvi97}.  However, their result contained diagrams in which the
spectator particle appeared off mass shell, clearly inconsistent with the
assumptions of the CST. The situation was even more confusing since
Adam and Van Orden \citep{Ada04} worked out an alternative derivation of the $3N$
current, using a purely algebraic technique, and obtained a result in which all
spectators are on mass shell. This apparent contradiction was resolved when
it was shown that the two results are in fact equivalent \citep{Gro04}. 

This equivalence is illustrated in Fig.\ \ref{F:equivalence}. Diagram (A)
is part of diagram (B), which has one spectator off mass shell. A closer
inspection shows that (A) is actually identical to (A') which in turn
contributes to (B') where all spectators are on mass shell. Applying this idea
one can replace all diagrams of type (B), such as the ones appearing in the
current of Kvinikhidze and Blankleider, by others satisfying the
spectator-on-mass-shell constraint.

\begin{figure}[t]
\epsfig{file=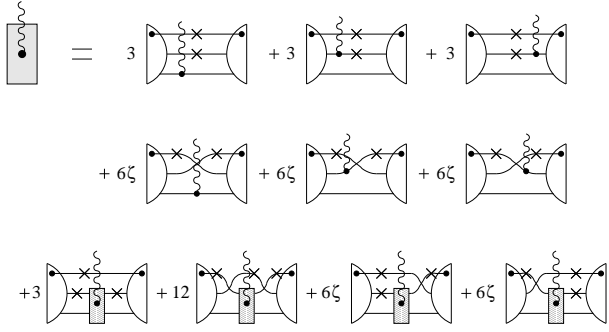,width=8cm}
\caption{The core diagrams of the electromagnetic $3N$ spectator current.} \label{F:core}
%
\end{figure}

\begin{figure}[b]
\epsfig{file=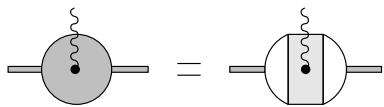,width=5cm}
\caption{The elastic electromagnetic spectator current, in terms of the core
diagrams.} \label{F:elcur}
%
\end{figure}

We derived the gauge invariant electromagnetic $3N$ current for the cases of
elastic and inelastic scattering from the $3N$ bound state within the spectator
theory in a diagrammatic approach \citep{Gro04}. The ``core diagrams'' of Fig.\
\ref{F:core} play a central role since they contribute in both cases. In fact,
elastic scattering (Fig.\ \ref{F:elcur}) is completely determined through the
core diagrams. The first 6 diagrams of Fig.\ \ref{F:elcur} comprise the Complete Impulse
Approximation (CIA) which is gauge invariant by itself. The remaining 4 diagrams involve the
photon coupling to the interacting $NN$ pair and belong therefore to the interaction
current.

\begin{figure}
\epsfig{file=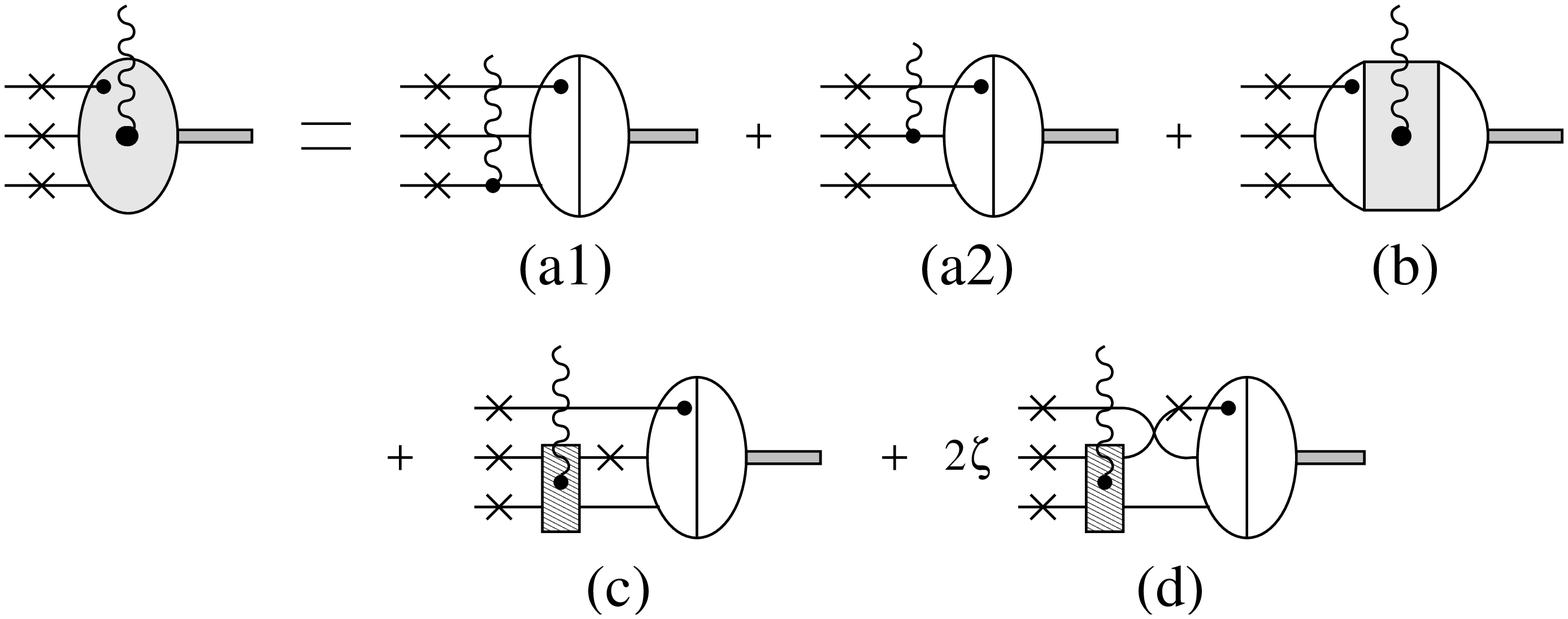,width=8cm}
\caption{The electromagnetic $3N$ breakup spectator current.} \label{F:breakupcur}
%
\end{figure}

In the case of electrodisintegration a number of diagrams need to be added in
which the photon couples to the outgoing nucleons. In Fig.\ \ref{F:breakupcur},
diagrams (a1) and (a2) represent the relativistic impuls approximation, (b)
contains the final state interactions, and (c) and (d) are interaction currents.

The derived currents will be used to calculate the elastic form factors of the $3N$ bound
state as well as its electrodisintegration. These calculations are currently in progress.

\begin{theacknowledgments}
This work was performed in collaboration with Franz Gross (Jefferson
Laboratory) and Teresa Pe\~na (IST Lisbon). It was supported by FEDER and the
{\em Funda\c c\~ao para a Ci\^encia e a Tecnologia} under grant
POCTI/FNU/40834/2001.
\end{theacknowledgments}

\end{document}